# Ultra Low Power 3D-Embedded Convolutional Neural Network Cube Based on $\alpha$-IGZO Nanosheet and Bi-Layer Resistive Memory

Sunanda Thunder[1], Parthasarathi Pal[2], Yeong-Her Wang[2]*, Po-Tsang Huang[1]*
[1]International College of Semiconductor Technology,
National Yang Ming Chiao Tung University, Hsinchu, Taiwan
[2]Institute of Microelectronics, National Cheng Kung University, Tainan, Taiwan
email: [1]bughuang@nctu.edu.tw , [2]yhw@ee.ncku.edu.tw

*Abstract*— In this paper we propose and evaluate the performance of a 3D-embedded neuromorphic computation block based on indium gallium zinc oxide ($\alpha$-IGZO) based nanosheet transistor and bi-layer resistive memory devices. We have fabricated bi-layer resistive random-access memory (RRAM) devices with $Ta_2O_5$ and $Al_2O_3$ layers. The device has been characterized and modeled. The compact models of RRAM and $\alpha$-IGZO based embedded nanosheet structures have been used to evaluate the system level performance of 8 vertically stacked $\alpha$-IGZO based nanosheet layers with RRAM for neuromorphic applications. The model considers the design space with uniform bit line (BL), select line (SL) and word line (WL) resistance. Finally, we have simulated the weighted sum operation with our proposed 8-layer stacked nanosheet based embedded memory and evaluated the performance for VGG-16 convolutional neural network (CNN) for *Fashion-MNIST* and CIFAR-10 data recognition, which yielded 92% and 75% accuracy respectively with drop out layers amid device variation.

## I. INTRODUCTION

The thousands of terabytes of data produced every second by trillions of internally and externally connected edge devices needs to be processed, filtered, and categorized. Therefore, the necessity of artificial intelligence has once again been reinvigorated in our daily life along with the need for data-centric computing to reduce the overheads at data centers. Software based artificial intelligence algorithm and neuromorphic system, implemented using traditional digital or mixed circuits, suffer from Von-Neumann bottleneck due to the physical separation between the computing unit and the memory, and the bandwidth limitation inhibits the performance of a powerful processor rendering it idle most of the time. Recent advents in the research of emerging non-volatile memories (eNVM) such as resistive random-access memory (RRAM), magnetic random-access memory (MRAM) and ferroelectric random-access memory (FeRAM) have accelerated the research of neuromorphic computing with an aim to alleviate memory bandwidth bottleneck in Von-Neumann computing architecture [1-12]. The invention of True-North chip by IBM [13], Loihi from Intel [14] tells the trend. Apart from this trend of building neuromorphic chips, the recent trend of monolithic 3D-IC design has also heavily impacted the IC design sector [15]. Although there have been many proposals for monolithic 3D-integration of neuromorphic chips, an integration of 3D-IC with eNVMs for neuromorphic applications is still missing.

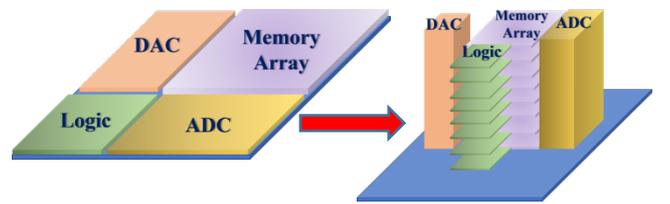

Fig.1 Stacked nanosheet based 3D computing in memory architecture provides energy, area and throughput efficiency over conventional CIM chip.

Among a many eNVMs resistive random-access RRAMs are one of the most promising candidates for neural network applications [16-21]. In this article we propose and demonstrate by simulation $\alpha$ -IGZO based novel 3D-embedded RRAM tile with common WL and SL for neuromorphic computing (Fig.1). The paper starts with a discussion of modeling the $\alpha$ -IGZO based nano-sheet transistors and RRAMs. We have found that $\alpha$-IGZO based nanosheet transistors are suitable for 3D-IC demonstration for neuromorphic computing. In the following section we discuss about the experiments and results, which is followed by discussion and conclusion.

## II. EXPERIMENTS AND RESULTS

The 3D-stacked $\alpha$-IGZO based nano-sheet transistors have been modeled according to [22]. The complete physical model ranging from subthreshold to strong inversion can be described as

$$I_D(j) = \frac{W}{L}\omega_0 \frac{T}{2T_e}\left(\frac{\epsilon_s}{C_i}\right)^{1-\frac{2T_e}{T}}\left\{\left[V_I(j)F\left(\frac{V_{GS}(j)}{V_I}\right)\right]^{\frac{2T_e}{T}}\right.$$
$$\left. - \left[V_I(j)F\left(\frac{V_{GSL}(j)}{V_I}\right)\right]^{\frac{2T_e}{T}}\right\}$$

$$I_{WL1} = \sum_{j=1}^{8} I_D(j)$$

Where, $V_{GS} = V_G - V_T - V_S$ , $V_{GSL} = V_G - V_T - V_{SL}$ , $f(x) = \log[1 + \exp(x)]$ and $V_I = V_i 2T_e/(2T_e - T)$.

Fig. 2(a). shows the schematic of a single column cell and Fig.2(b) and Fig.2(c) shows the drain current from a single

nano-sheet layer for linear and saturation region. We have fabricated, characterized, and modeled bi-layer RRAM with 5 nm $Ta_2O_5$ and 2 nm $Al_2O_3$. Fig. 2(d). shows the I-V curve of resistive random-access memory, which has been modeled according to [23]. The pulse program erase operation in a single RRAM device showed 32 distinct states (not shown here), however in presence of device variation with σ=0.1, 32 states merge and only 3 distinct states are available.

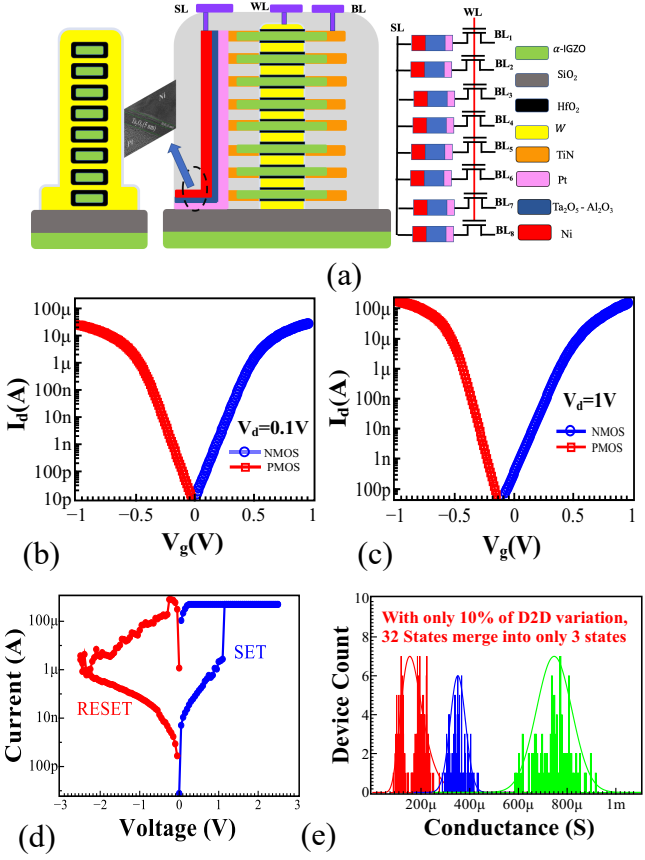

(a)
(b)
(c)
(d)
(e)

Fig. 2 (a). The schematic of the fabricated and proposed device with transmission electron microscopic (TEM) image. (b) The $I_d$-$V_g$ curve for α-IGZO based nanosheet transistors in linear region with $V_{ds}$=0.1 V. (c) The $I_d$-$V_g$ curve for α-IGZO based nanosheet transistors in saturation region with $V_{ds}$=1 V. (d). The I-V curve for bi-layer RRAM devices considered in this study. (e). The device to device variation in RRAM.

After the characterization and modeling of a single column cell, we have simulated the multiply-accumulate operation in a single tile. The design space of a single tile has been considered according to [24]. The BL, WL and SL resistance has been calculated according to [25]. We have considered a single tile with 8 × 8 column cells and each column cell consists of 8 nanosheet transistors. The 3D-integration of RRAM with nanosheet provides us with an edge over 8x area efficiency. Fig.3(a). shows the architecture of a single tile and fig.3(b) shows the MAC ops output. The pivotal part of the MAC core is the 3D-memory array and the peripheral circuits. The peripherals consists of select cell coding system (SCC), data and level converters, analog digital converters (ADCs), digital analog converters (DACs), adders and shift registers, and rectified linear unit (ReLu) activation function. We have simulated the MAC ops accroding to our previous modeling and assumptions on WL, BL and SL resistance. During MAC operation the static current of a single column cell is in nA range, which facilitates ultra-low power operation in a single tile. This tile based architecture has is the key to avoid IR drop across the WL, SL and BL along with this this tile based architecture also facilitates the implementation of drop out mechanism in convolutional neural network, which has been discussed in the following section.

To facilitate low power operation, *"READ"* operation has been conducted by applying on 0.5 V. The maximum "ON" current is 10 μA for 2V *"WRITE"* and 0.5V *"READ"*. The *"WRITE"* operation has been conducted by sending pulses through switch matrix circuit depicted in [25]. The *"READ"* operation has performed by charging a capacitor along with a high resolution 4-bit successive approximation analog to digital converter (SAR-ADC). The 4-bit ADC reduces the complexity of the peripheral circuit and reduces the power consumption.

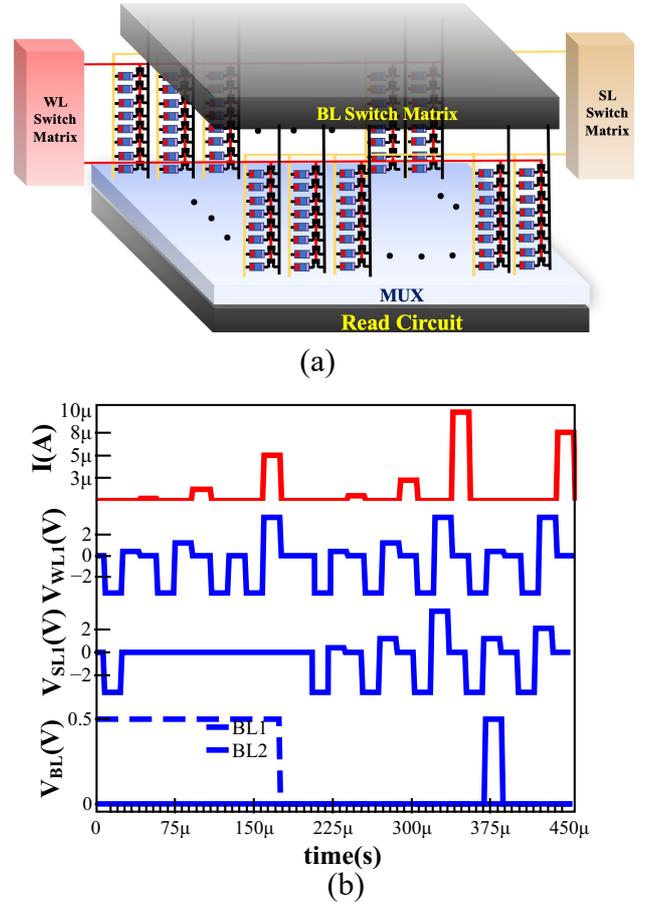

(a)
(b)

Fig. 3 (a). The schematic of the proposed 3d-embedded convolutional neural network cube based on α-IGZO nanosheet and bi-layer resistive memory devices. (b). The simulated MAC operation output from the experimentally calibrated data.

Following the circuit implementation of a single tile we have used this same tile-based network for evaluating the performance of convolutional neural network for recognizing the images from the F-MNIST and CIFAR-10 data set. We have used VGG-16 model here, which has been depicted in Fig.4(a). *Rectified Linear Unit* activation function has been used to each layer so that all the negative values are not passed to the next layer. Fig.4(a). shows the architecture of the CNN used in this work for evaluating the system level performance of the proposed 3D-synaptic device. The tile-based architecture is shown in fig.4(b). The tile-based network serves two different purposes in this works.

1. Alleviation of design space IR- drop across BL, WL and SL.
2. Implementing drop out mechanism to avoid over fitting in training data.

The drop out is also an important mechanism in CNN, which helps avoiding the overfitting of the data. Dropout is a typical and simple mechanism that will randomly drop nodes from the network during training process. This is used to avoid over-fitting during training operation. In this work instead of dropping out nodes we drop out and entire column network from a single tile. During the training process, a pseudo random numbers were generated select the drop-out column in each tile.

Prior to drop out the training accuracy for FMNIST was 98%, whereas the inference accuracy was only 91%. Post drop-out the training accuracy was 93% and inference accuracy is 92%. The CIFAR-10 training accuracy was 81% and 75% inference accuracy has been achieved. Fig. 4(c) shows the inference accuracy for F-MNIST data set and fig.4(d) shows the inference accuracy for CIFAR-10 data set.

simulation shows that the proposed architecture can detect Fashion-MNIST data set with 92% accuracy. (d). (c). Neuromorphic simulation shows that the proposed architecture can detect CIFAR-10 data set with 75% accuracy.

The tile-array-based accelerator performs VGG-16 CNN. Compared to a digital accelerator with the same quantitation (8b/4b for activation/weight) and channel pruning, the accuracy for F-MNIST data recognition is 92% and CIFAR-10 data is 75%. Table-I performs the system level bench marking of this work with other state of art works.

|  | This Work | [26] | [12] |
|---|---|---|---|
| # States | 3 | 2 | 27 |
| $\sigma_{D2D}$ | 10% | 5% | 0.57833V |
| $R_{on}/R_{off}$ | $10^6$ | $4.2 \times 10^3$ | $10^5$ |
| Dataset | CIFAR-10/ FMNIST | MNIST | MNIST |
| Accuracy | 75% / 92% | 97.11% | 96% |

### III. CONCLUSION

The newly proposed $\alpha$-IGZO based 3D memory was demonstrated to exhibit superior performance with high speed and incremental switching. According to the excellent electrical characteristics, a mini-array-based accelerator was further demonstrated with CONV-Net for objection detection. Instead of network-level weight stationary, this accelerator utilized layer-level weight stationary to reduce both area and energy by using 3D stacking. Moreover, standard or pointwise convolutions and channel pruning were supported by the channel-based weight mapping scheme with an accuracy of 92% for F-MNIST data set and 75% accuracy with CIFAR-10 dataset.

### IV. ACKNOWLEDGEMENT

This work was financially supported by the "Center for the Semiconductor Technology Research" by the Ministry of Education (MOE) in Taiwan. Also supported in part by the Ministry of Science and Technology, under Grant MOST 110-2634-F-009-027, MOST 110-2218-E-A49-014-MBK and MOST-108-2221-E-006-040-MY3. We are grateful to Center for Micro/Nano Science and Technology, NCKU for providing us with experimental platform. We are also grateful to Dr. Sourav De for providing us with important insights on CNN and 3D-IC design.

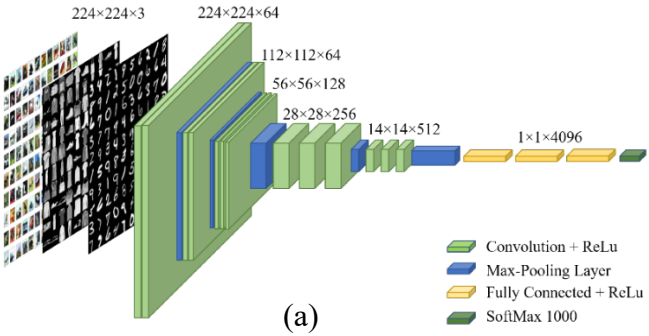
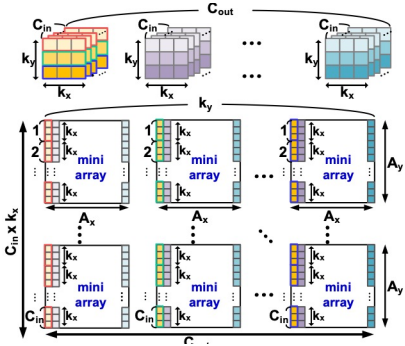
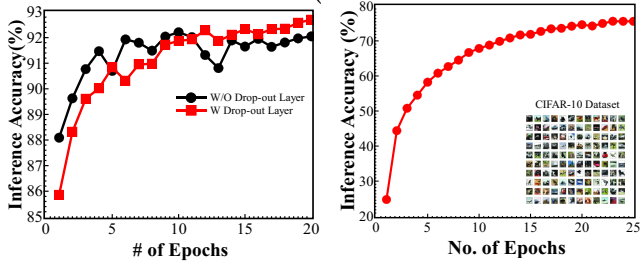

Fig.4. (a). Architecture of CNN considered during neuromorphic simulation. (b) The tile-based architecture considered in this work to minimize the I-R drop across the BL, WL and SL. (c). Neuromorphic

### V. REFERENCE

[1] Sourav De, Darsen Lu, et al.,"*Ultra-Low Power Robust 3bit/cell Hf$_{0.5}$Zr$_{0.5}$O$_2$ Ferroelectric FinFET with High Endurance for Advanced Computing-In-Memory Technology*". In: 2021 Symposia on VLSI Technology and Circuits. 2021.

[2] Chen, WH., Dou, C., Li, KX. *et al. CMOS-integrated memristive non-volatile computing-in-memory for AI edge processors.* Nat Electron **2,** 420–428 (2019).

[3] M. Jerry et al., "*Ferroelectric FET analog synapse for acceleration of deep neural network training,*" in 2017 IEEE International Electron Devices Meeting (IEDM), San Francisco, CA, USA, 2017, pp. 6.2.1–6.2.4, doi: 10.1109/IEDM.2017.8268338.

[4] Gong, N., Idé, T., Kim, S. *et al.* *Signal and noise extraction from analog memory elements for neuromorphic computing. Nat Commun* **9,** 2102 (2018).

[5] Darsen Duane Lu, Sourav De, *et al.,* "*Computationally efficient compact model for ferroelectric field-effect transistors to simulate the online training of neural networks,*" *Semiconductor Science and Technology*, vol. 35, 095007, 2020.

[6] Sourav De, *et al.,* "*Tri-Gate Ferroelectric FET Characterization and Modelling for Online Training of Neural Networks at Room Temperature and 233K,*" *2020 Device Research Conference (DRC)*, 2020, pp. 1-2, doi: 10.1109/DRC50226.2020.9135186.

[7] C. Li et al., "*A Scalable Design of Multi-Bit Ferroelectric Content Addressable Memory for Data-Centric Computing,*" 2020 IEEE International Electron Devices Meeting (IEDM), 2020, pp. 29.3.1-29.3.4, doi: 10.1109/IEDM13553.2020.9372119. S. De et al., "*Robust Binary Neural Network Operation from 233 K to 398 K via Gate Stack and Bias Optimization of Ferroelectric FinFET Synapses,*" in IEEE Electron Device Letters, doi: 10.1109/LED.2021.3089621.

[8] J. -Y. Ciou, S. De, C. -W. Wang, W. Lin, Y. -J. Lee and D. Lu, "*Analytical Modelling of Ferroelectricity Instigated Enhanced Electrostatic Control in Short-Channel FinFETs,*" 2021 5th IEEE Electron Devices Technology & Manufacturing Conference (EDTM), 2021, pp. 1-3, doi: 10.1109/EDTM50988.2021.9420931.

[9] Sourav De *et al.* "*Neuromorphic Computing with Deeply Scaled Ferroelectric FinFET in Presence of Process Variation, Device Aging and Flicker Noise*", arXiv:2103.13302v1.

[10] Sourav De, et al., "*Uniform Crystal Formation and Electrical Variability Reduction in Hafnium-Oxide-Based Ferroelectric Memory by Thermal Engineering*", *ACS Applied Electronic Materials* **2021** *3* (2), 619-628, DOI: 10.1021/acsaelm.0c00610.

[11] Sourav De, *et al.,* "*Alleviation of Charge Trapping and Flicker Noise in HfZrO2-Based Ferroelectric Capacitors by Thermal Engineering,*" 2021 International Symposium on VLSI Technology, Systems and Applications (VLSI-TSA), 2021, pp. 1-2, doi: 10.1109/VLSI-TSA51926.2021.9440091.

[12] Sourav De, *et al.,* "*Stochastic Variations in Nanoscale HZO based Ferroelectric finFETs: A Synergistic Approach of READ Optimization and Hybrid Precision Mixed Signal WRITE Operation to Mitigate the Implications on DNN Applications.*", arXiv:2008.10363v3.

[13] F. Akopyan et al., "*TrueNorth: Design and Tool Flow of a 65 mW 1 Million Neuron Programmable Neurosynaptic Chip,*" in IEEE Transactions on Computer-Aided Design of Integrated Circuits and Systems, vol. 34, no. 10, pp. 1537-1557, Oct. 2015, doi: 10.1109/TCAD.2015.2474396

[14] M. Davies et al., "*Loihi: A Neuromorphic Manycore Processor with On-Chip Learning,*" in IEEE Micro, vol. 38, no. 1, pp. 82-99, January/February 2018, doi: 10.1109/MM.2018.112130359.

[15] S. . -W. Chang *et al*., "*First Demonstration of CMOS Inverter and 6T-SRAM Based on GAA CFETs Structure for 3D-IC Applications,*" *2019 IEEE International Electron Devices Meeting (IEDM)*, 2019, pp. 11.7.1-11.7.4, doi: 10.1109/IEDM19573.2019.8993525.

[16] Chen, YC., Lin, CC., Hu, ST. *et al. A Novel Resistive Switching Identification Method through Relaxation Characteristics for Sneak-path-constrained Selector less RRAM application. Sci Rep* **9,** 12420 (2019).

[17] D. Ielmini, F. Nardi, and C. Cagli, "*Resistance-dependent amplitude of random telegraph-signal noise in resistive switching memories*" Appl. Phys. Lett. vol. 96, 2010, 053503.

[18] D. Veksler, G. Bersuker, B. Chakrabarti, E. Vogel, S. Deora, K. Matthews, D. C. Gilmer, H.-F. Li, S.Gausepohl, P. D. Kirsch "*Methodology for the statistical evaluation of the effect of random telegraph noise (RTN) on RRAM characteristics*",In IEEE International Electron Devices Meeting (IEDM), 2012.

[19] G. Bersuker, DC Gilmer, D Veksler, J Yum, H Park, S Lian, L Vandelli, A, Padovani, L. Larcher, K. McKenna, A. Shluger, V. Iglesias, M. Porti,M. Nafria,"*Metal oxide RRAM switching mechanism based on conductive filament microscopic properties*"Journal Appl. Phys., vol.110, 2011, p. 124518.

[20] L. Vandelli, A. Padovani, L. Larcher, G. Broglia, G. Ori, M. Montorsi, G. Bersuker, and P. Pavan, "*Comprehensive physical modeling of forming and switching operations in HfO2 RRAM devices,*" IEEE International Electron Devices Meeting (IEDM) Tech. Digest, 2011, pp. 421-424.

[21] Parthasarathi Pal *and* Yeong Her Wang , "*Interconversion of complementary resistive switching from graphene oxide based bipolar multilevel resistive switching device*", Applied Physics Letters 117, 054101 (2020) https://doi.org/10.1063/5.0010319

[22] M. Ghittorelli et al., "*Compact physical model of a-IGZO TFTs for circuit simulation,*" 2017 47th European Solid-State Device Research Conference (ESSDERC), 2017, pp. 98-101, doi: 10.1109/ESSDERC.2017.8066601.

[23] X. Guan, S. Yu and H.-S. P. Wong*,* "*A SPICE compact model of metal oxide resistive switching memory with variations,*" IEEE Electron Device Letters, vol. 33, no. 10, pp. 1405-1407, 2012.

[24] T. Yang *et al*., "*Design Space Exploration of Neural Network Activation Function Circuits,*" in *IEEE Transactions on Computer-Aided Design of Integrated Circuits and Systems*, vol. 38, no. 10, pp. 1974-1978, Oct. 2019, doi: 10.1109/TCAD.2018.2871198.

[25] P. Chen, X. Peng and S. Yu, "*NeuroSim+: An integrated device-to-algorithm framework for benchmarking synaptic devices and array architectures,*" *2017 IEEE International Electron Devices Meeting (IEDM)*, 2017, pp. 6.1.1-6.1.4, doi: 10.1109/IEDM.2017.8268337.

[26] P. Pal, S. Thunder, M. -J. Tsai, P. -T. Huang and Y. H. Wang, "*Benchmarking the Performance of Heterogeneous Stacked RRAM with CFETSRAM and MRAM for Deep Neural Network Application Amidst Variation and Noise,*" *2021 International Symposium on VLSI Technology, Systems and Applications (VLSI-TSA)*, 2021, pp. 1-2, doi: 10.1109/VLSI-TSA51926.2021.9440130.